\documentclass{aa}
\usepackage{graphicx}
\usepackage[varg]{txfonts}
\usepackage{natbib}        	
\usepackage{color}      	
\usepackage{hyperref}

\newcommand{\frqa}{327.0}
\newcommand{\frqb}{410.5}
\newcommand{\frqc}{432.0}
\newcommand{\frqaM}{327.0\,MHz}
\newcommand{\frqbM}{410.5\,MHz}
\newcommand{\frqcM}{432.0\,MHz}

\newcommand{\kms}{\,km\,s$^{-1}$}
\newcommand{\etal}{et al. }
\newcommand{\Alfven} {Alfv\'{e}n}
\newcommand{\be}{\begin{equation}}
\newcommand{\ee}{\end{equation}}
\newcommand{\beq}{\begin{eqnarray}}
\newcommand{\eeq}{\end{eqnarray}}
\graphicspath{ {./Fig_Fibers_NRH/}}
\begin{document}
 \title{High resolution observations with ARTEMIS-JLS and the NRH}
 \subtitle{(IIb) Spectroscopy and imaging of fiber bursts}

   \author{C.E. Alissandrakis \inst{1}
              C. Bouratzis \inst{2}
   \and
   		A. Hillaris \inst{2}
}
\offprints{C.E. Alissandrakis}
\institute{			Department of Physics, University of Ioannina, 45110 Ioannina, Greece\\
\email{calissan@cc.uoi.gr}
\and				Department of Physics, University of Athens, 15783 Athens, Greece
}
\authorrunning{Alissandrakis et al.}
\titlerunning{Imaging of fiber bursts}
 \date{Received .....; accepted ......}
\abstract
{}
{We study the characteristics of intermediate drift bursts (fibers) embedded in a large type IV event.}
{We used high sensitivity, low noise dynamic spectra obtained with the acousto-optic analyzer (SAO) of the ARTEMIS-JLS solar radiospectrograph, in conjunction with high time resolution images from the Nan\c cay Radioheliograph (NRH) and EUV images from TRACE to study fiber bursts during the July 14, 2000 large solar event. We computed both 2-dimensional and 1-dimensional images and applied high pass time filtering to the images and the dynamic spectrum in order to enhance the fiber-associated emission. For the study of the background continuum emission we used images averaged over several seconds.}
{Practically all fibers visible in the SAO dynamic spectra are identifiable in the NRH images. Fibers were first detected after the primary energy release in a moving type IV,  probably associated with the rapid eastward expansion of the flare and the post-flare loop arcade. We found that fibers appeared as a modulation of the continuum intensity with a root mean square value of the order of 10\%. Both the fibers and the continuum were strongly circularly polarized in the ordinary mode sense, indicating plasma emission at the fundamental. We detected a number of discrete fiber emission sources along two $\sim300$\,Mm long parallel stripes, apparently segments of large scale loops encompassing both the EUV loops and the CME-associated flux rope. We found cases of multiple fiber emissions appearing at slightly different positions and times; their consecutive appearance can give the impression of apparent motion with supra-luminal velocities. Images of individual fibers were very similar at \frqc\ and \frqaM. From the position shift of the sources and the time delays at low and high frequencies we estimated,  for a well observed group of fibers, the exciter speed and the frequency scale length along the loops; we obtained consistent values from imaging and spectral data, supporting the whistler origin of the fiber emission. Finally we found that fibers in absorption and in emission are very similar, thus confirming that they are manifestations of the same wave train.}
{}

   \maketitle

\section{Introduction}\label{intro}

Fibers are solar radio bursts with intermediate frequency drift  ({\it i.e.} between the slow drifting type II burst and the fast drifting type III bursts);  they are observed in groups at decimetric and lower frequencies, embedded in type IV burst continua  \citep{Young61, Elgaroy73, Slottje1972, Fomichev78, Slottje1981}. In addition to their filamentary shape in dynamic spectra, another important characteristic is that they appear in the form of absorption/emission ridges, the absorption having a slightly lower frequency than the emission.

Intermediate drift bursts are believed to be signatures of an exciter moving along post-flare magnetic loops. The exciter is most likely a train of whistler waves, radiating through their interaction with Langmuir waves \citep{Kuijpers1972, Kuijpers1975, Kuijpers80}. According to this interpretation, the emission is enhanced at $\rm \omega_{pe} + \omega_w$ and reduced at $\rm \omega_{pe}$ where $\rm \omega_w$ is the whistler frequency and $\rm \omega_{pe}$ the plasma frequency. The frequency range of the emission is limited by the condition $\rm 0.25 \le \omega_w/ \omega_{ce} \le 0.5$, where $\rm \omega_{ce}$ is the electron cyclotron frequency; the first part of this relation is the necessary condition for the development of the instability and the second for strong cyclotron damping. Alternative interpretations involve \Alfven--Langmuir wave  interaction \citep{Bernold83}, a combination of \Alfven~waves and electron cyclotron maser \citep{Benz98}, or  modulation of the background type IV continuum by fast magneto-acoustic wave trains \citep{Kuznetsov2006, Karlicky2013, zlobec2014}. Whatever their origin, the intermediate drift bursts qualify as coronal magnetic field diagnostics \citep[see~][]{2005A&A...435.1137A, 2007SoPh..245..327R, Rausche08, Bouratzis2019}.

Combined analysis of dynamic spectra and images has been done in the past by  \cite{2005A&A...435.1137A}, who used 2D  positions from the Nan\c cay Radioheliograph, together with potential extrapolations of the photospheric magnetic field and an $a\times$ Newkirk density model to identify the magnetic loops in which fiber bursts occurred. Their best fit was for $a=3.5$ and they deduced field strengths from 6-14\,G at 410\,MHz (height of 20\,Mm) to 3\,G higher up, at 100\,Mm (236\,MHz). A similar analysis was performed by \cite{2007SoPh..245..327R}. Recently, \cite{2017ApJ...848...77W} studied three groups of fiber bursts in the 1--2 GHz range; they used dynamic imaging spectroscopy with the Karl G. Jansky Very Large Array (VLA) which allowed the reconstruction of spatially resolved fiber trajectories. The comparison of their results with theoretical models favors the whistler wave model and points to the origin of the fiber bursts at or near the footpoints of coronal loops.

\begin{figure*}[t]
\hspace{3cm}\hspace{0.77cm}
\includegraphics[width=.588\textwidth]{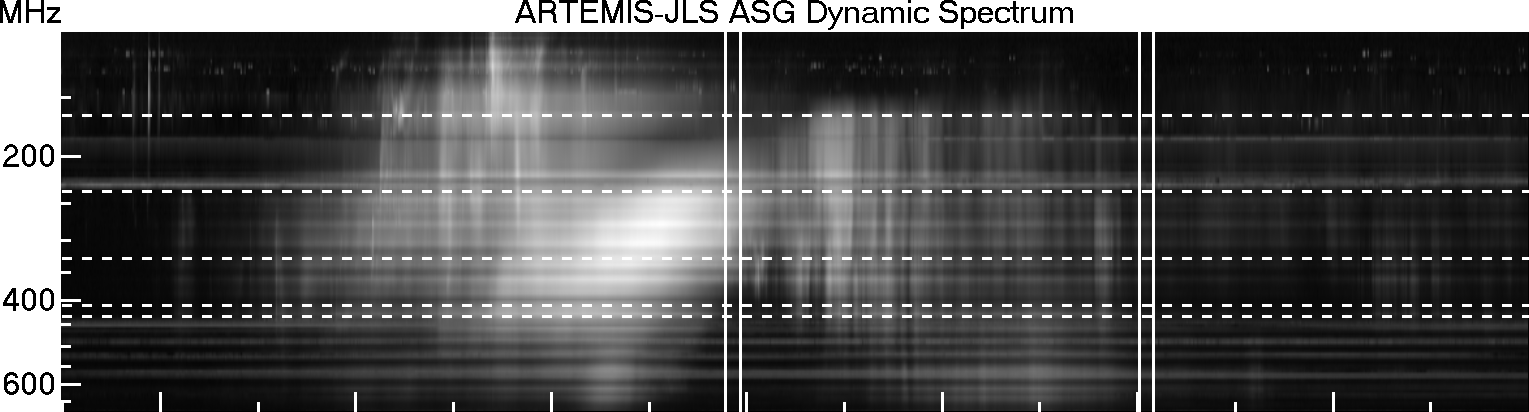}

\medskip
\hspace{3cm}
\includegraphics[width=.65\textwidth]{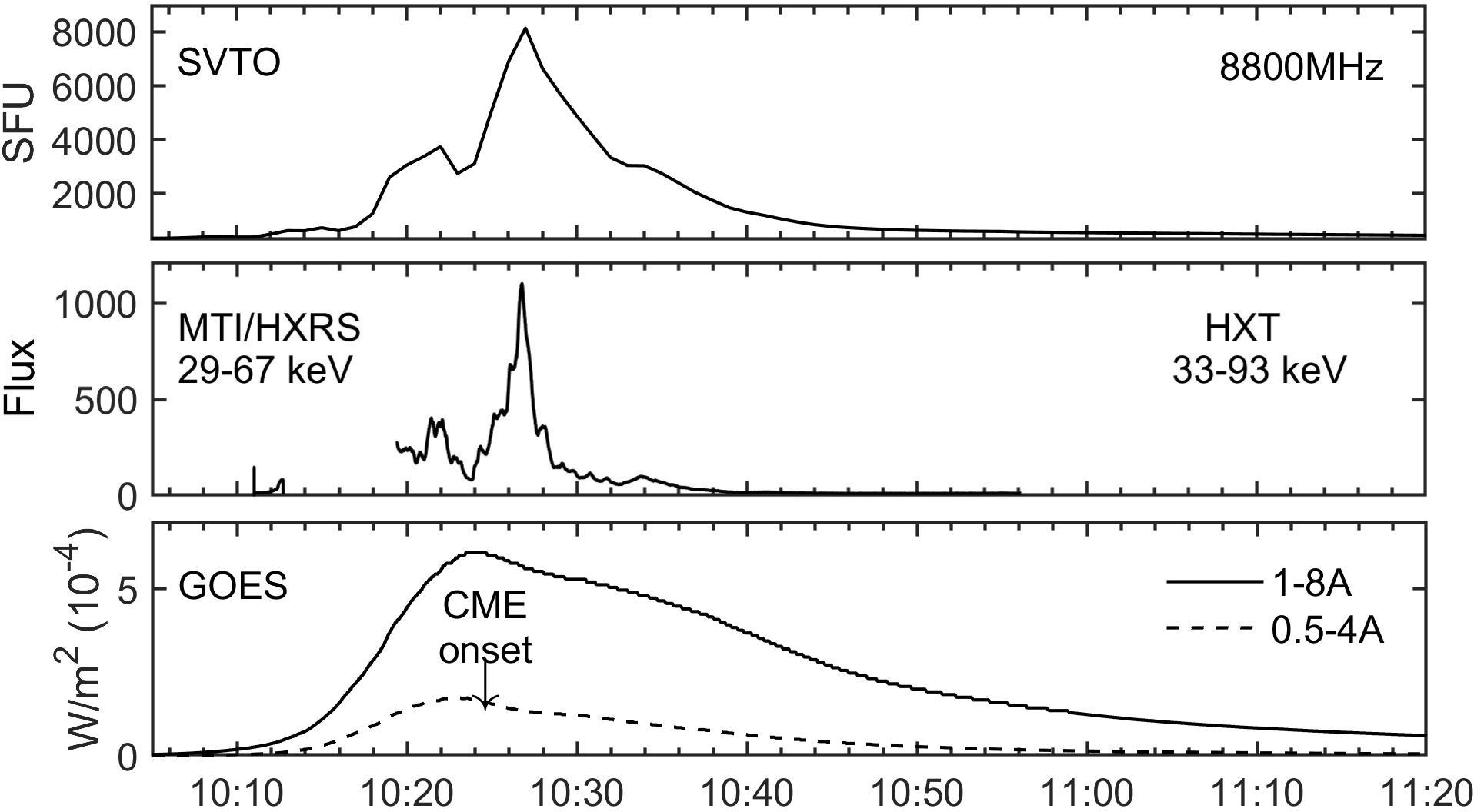}
\caption{Dynamic spectrum of the July 14, 2000 event recorded by the ARTEMIS-JLS sweep frequency receiver, ASG, in the 100--670\,MHz range. Dashed horizontal lines mark the NRH frequencies, vertical lines the two selected intervals of fiber emissions. The bottom plots show time curves of microwave emission at 2695 MHz from San Vito, 33-93\, keV hard X-rays form {\it the Hard X-ray Spectrometer} (HXRS) on board the {\it Multi-Spectral Thermal Imager} (MTI) and the {\it Hard X-ray Telescope} (HXT) on board Yohkoh and soft X-rays from GOES.}
\label{FullASG}
\end{figure*}

In a recent work (Bouratzis et al. 2019, hereafter paper I) we performed an extensive study of a large number of fibers embedded in sixteen type IV bursts, observed with the high sensitivity, 10\,ms resolution receiver of the ARTEMIS-/Jean-Louis Steinberd (JLS) radio spectrograph. We distinguished six morphological groups of fibers and measured their duration, their bandwidth, and drift rate as a function of time. We found that the whistler hypothesis leads to reasonable magnetic field values, while the \Alfven\ origin required too high a magnetic field; we derived an average magnetic field of 4.4\,G and an average frequency scale length along the loop of 220\,Mm.    

In this work we extend the work presented in paper I by using combined ARTEMIS-JLS and Nan\c cay Radioheliograph (NRH) to study fibers during the July 14, 2000 flare. In Sect \ref{obs} we present the observations and in Sect. \ref{results} our results. We conclude with the discussion in Sect. \ref{conclusions}

\section{Observations and their analysis}\label{obs}
The observations with ARTEMIS-JLS \citep{Caroubalos01} and their analysis were described in detail in paper I. Here we only mention that the fibers were observed with the acousto-optic analyzer (SAO) in the frequency range of 265--470\,MHz with a spectral sampling of 1.4\,MHz and a temporal resolution of 10\,ms. For the overview of the event we used data from the sweep-frequency Global Spectral Analyser (ASG), which operated in the 100-700\,MHz range with a time resolution of 100\,ms.

The Nan\c cay Radioheliograph \citep{Kerdraon97} is a synthesis instrument that provides two-dimensional images of the Sun with sub-second time resolution. For the event that we study here, the NRH provided data at five frequencies (164.0, 236.6, 327.0, 410.5 and 432.0 MHz) with a cadence of 125 ms. All five frequencies are within the spectral range of the ASG, while the last three are also within the range of the SAO. 

From the original NRH visibilities, we computed two-dimensional (2D) images with a resolution of 1.04\arcmin\ by 1.52\arcmin\ at 432 MHz. In a previous work on spike bursts \citep{Bouratzis2016} we had also computed one-dimensional (1D) images using visibilities from the east-west (EW) and north-south (NS) arrays only; this improved the resolution by a factor of 2, due to the fact that the extension antennas make a very small contribution to the 2D images. In the present case these 1D images were rather noisy, whereas the fiber sources were well resolved and thus we preferred to compute 1D images by integrating the 2D images along the EW and NS directions. This gave us a less noisy and more robust result; no detailed self-calibration was necessary, but we did delete some defective baselines.

\section{Results}\label{results}

\subsection{Overview of the event of July 14, 2000}
This is a famous event described by many authors \citep[e.g.][]{2001A&A...373.1073K, 2001SoPh..204...55M, 2001SoPh..204..153W}. ARTEMIS data were presented by \cite{2001SoPh..204..165C}, see also Fig. 17 of \cite{2015SoPh..290..219B}. It was a complex event that occurred in active region 9077 and was associated to an X5.7 flare, starting at 10:03 UT near the center of the disk, and to a halo CME. The flare was quite extended in longitude (from $\sim6$\degr\,East to $\sim8$\degr\,West, latitude $\sim16$\degr), with spectacular loops recorded at 195\,\AA\ by the {\it Transition Region and Coronal Explorer} (TRACE).

In the early phase of the event, which included the filament activation and the CME onset, the energy release was limited in the west part of active region \citep{2001A&A...373.1073K, 2001SoPh..204..165C}. The expansion of the rising filament was also directed west of the flare, and so was the bulk of the CME, with a measured position angle of 273\degr. Near the Hard X-ray (HXR) peak at 10:26:45 UT, the flare emission expanded rapidly towards the east part of the AR and post-flare loops formed all along the neutral line.  

The metric emission (Fig. \ref{FullASG}) was dominated by a drifting continuum that started after the GOES and HXR peaks, around 10:27 UT and lasted up to about 10:41 UT, in the frequency range 200-450\,MHz. We also have a lower intensity broad-band continuum throughout the event, numerous IIIs and a faint type II. The continuum was rich in embedded pulsations, spikes, fibers and zebra patterns. Fiber bursts appeared around  10:33 UT and lasted until about 11:25 \citep[Table 2 of ][]{2015SoPh..290..219B}, often mixed with pulsations and other broadband structures \citep[see Figs 6-7 in ][]{2001SoPh..204..165C}. Their bulk parameters during the drifting continuum are given in entry 11 of Table A.1 of paper I and in entry 12 for the time interval after that.

Although some strong fibers are also visible in the ASG spectra, none was detected at the NRH frequency of 236\,MHz; we thus limited our comparative study to the three NRH frequencies that fall within the SAO range (\frqaM. \frqbM\ and \frqcM). We selected two 45\,s long time intervals, in which fiber bursts were relatively uncontaminated by other emissions; these are marked by vertical lines in the dynamic spectrum of Fig. \ref{FullASG}. The first interval, from 10:38:55 to 10:39:40 UT is near the end of the drifting continuum emission and the second interval, from 11:00:05 to 11:00:50 UT, is well after that, when the continuum emission was rather weak.

\subsection{Evolution of the continuum source}\label{ContEvol}
\begin{figure}
\begin{center}
\includegraphics[width=\hsize]{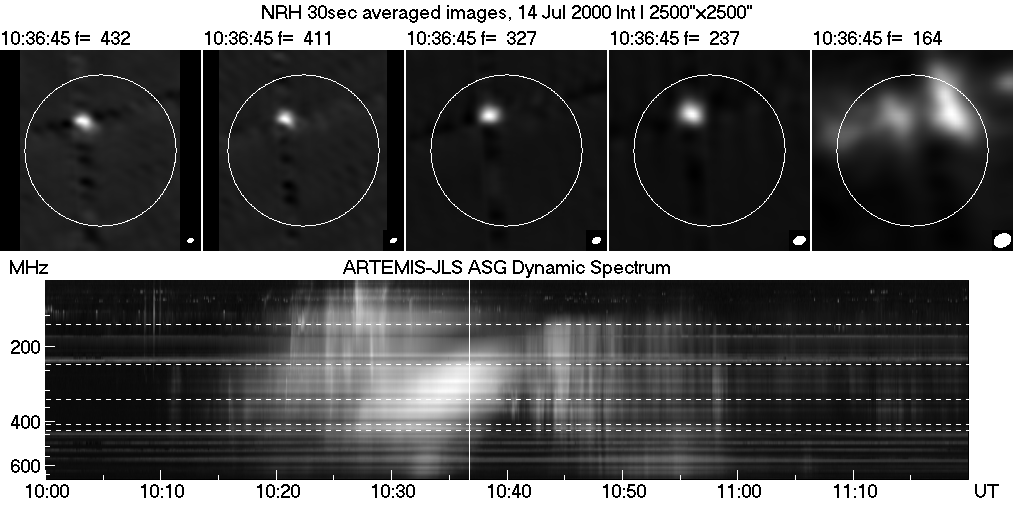}
\end{center}
\caption{A frame from Movie 1, showing 30-sec average NRH images in Stokes I at all five frequencies. The images have been normalized so that the minimum intensity of each image is black and the peak intensity is white. The white circles mark the photospheric limb. The NRH beam is drawn in the lower right corner of each frame. The bottom panel shows the ASG dynamic spectrum, with the vertical white line marking the time of the images and the dashed horizontal lines marking the NRH frequencies.}
\label{30secAver}
\end{figure}

Before we proceed with the fibers, it is interesting to discuss the evolution of the continuum source. For this purpose we will use NRH images with a 30\,s time integration. We note that the modulation of the intensity due to the fibers was small, $\sim10$\% in rms, thus the 30\,s average images are representative of the background continuum. Fig. \ref{30secAver} and the associated Movie 1 show the evolution of the radio sources at the five NRH frequencies, together with the ASG dynamic spectrum for reference. In order to compensate for the strong intensity variability, we have normalized each NRH image by its maximum value. 

\begin{figure}[h]
\begin{center}
\includegraphics[width=.6\hsize]{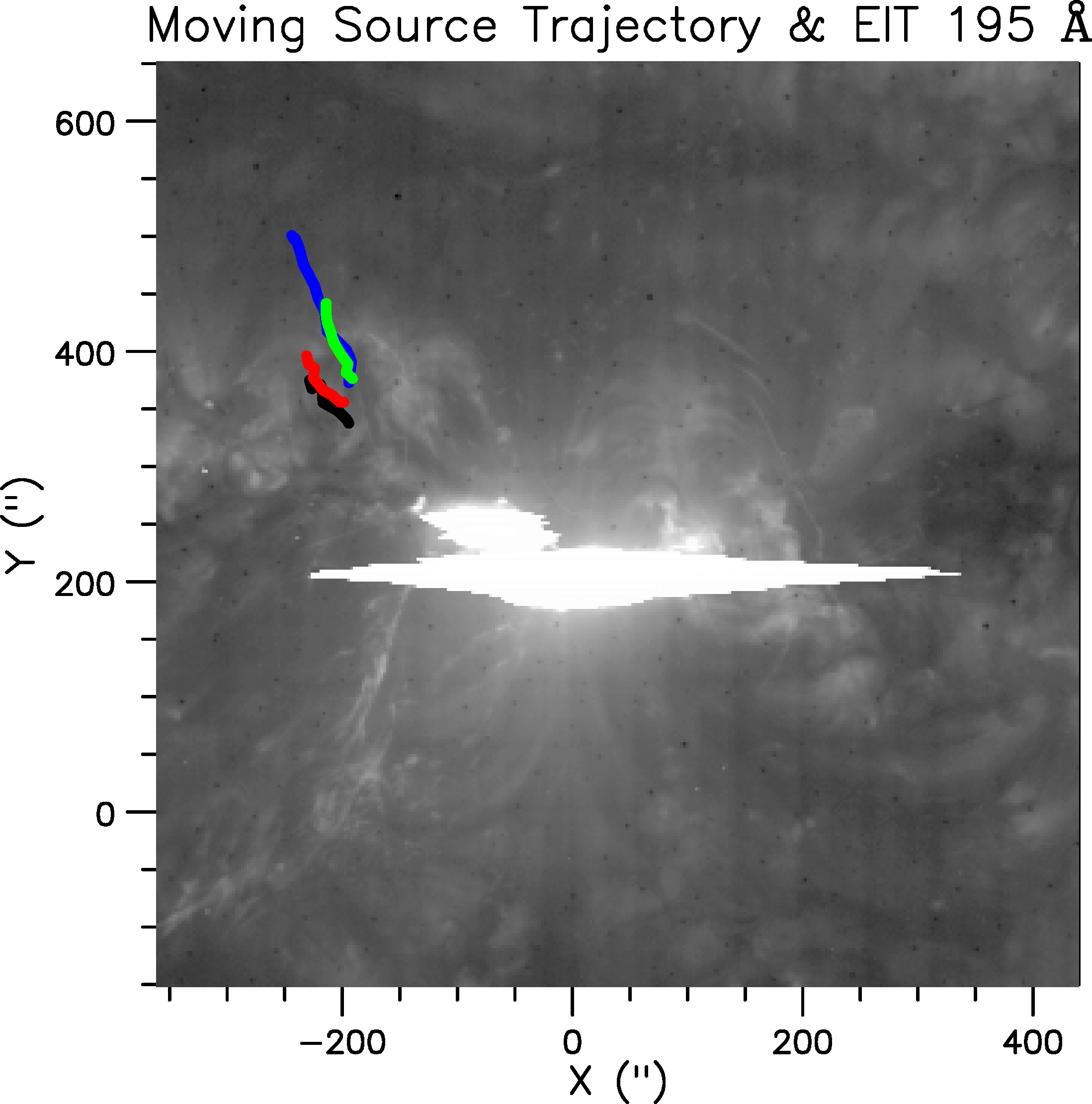}
\end{center}
\caption{The trajectories of the moving source at \frqcM\ (black), \frqbM\ (red), \frqaM\ (green) and 236.6\,MHz (blue) on top of an EIT 195\,\AA\ image at 10:36:10 UT. At \frqc, \frqb\ and \frqaM\ the trajectories are from 10:33:15 to 10:36:45 UT and at 236.6\,MHz from 10:33:15 to 10:39:15 UT}
\label{moving}
\end{figure}

During the early phase of the event, most of the metric emission came from the North-West quadrant of the solar disk, apparently associated with the primary energy release and the beginning of the halo CME, as discussed in the previous section and described by \cite{2001A&A...373.1073K} and \cite{2001SoPh..204..165C}. The movie frame of Fig. \ref{30secAver} shows emission that appeared North-East of the flare \citep[eastern source in][]{2001A&A...373.1073K} from around 10:27 UT at high frequencies. From 10:34 UT this source dominated the drifting continuum and the entire dynamic spectrum above 164.0\,MHz, while from 10:43 UT it was the only source at 164.0\,MHz as well. Most important for us here, it was the source of the first fibers. 

From about 10:33 UT, this source displayed a clear rectilinear motion to the North-East, at all frequencies except for 164\,MHz. The motion is well visible in Movie 1; at the high NRH frequencies it lasted until 10:37 UT, and until 10:39 UT at 236.6\,MHz. Although after that it is difficult to follow the motion of the continuum source due to the presence of broad-band structures (Fig. \ref{FullASG}), there was a development of multiple components and a net displacement to the east and south, indicating a shift of the emission source lower in the corona. At 164\,MHz which, as noted above, was outside the drifting continuum, the emission did not follow the same evolution.

Fig. \ref{moving} shows the trajectories of the source during the upward rectilinear motion, overplotted on an EIT image at 195\,\AA; although the flare image is saturated, it is clear that the extensions of the trajectories cross the eastern edge of the flare-loop arcade. The projected distance from that ranged from $\sim$140\arcsec\ for the \frqcM\ source at the start of the rectilinear motion, to $\sim$300\arcsec\ for the \frqaM\ source at the end, which gives an indication of the height. We note that the position and the direction of motion at \frqcM\ and \frqbM\ was slightly different than in the other two frequencies, which might indicate that we see two different magnetic structures. 

\begin{figure*}[t]
\begin{center}
\includegraphics[width=.8\hsize]{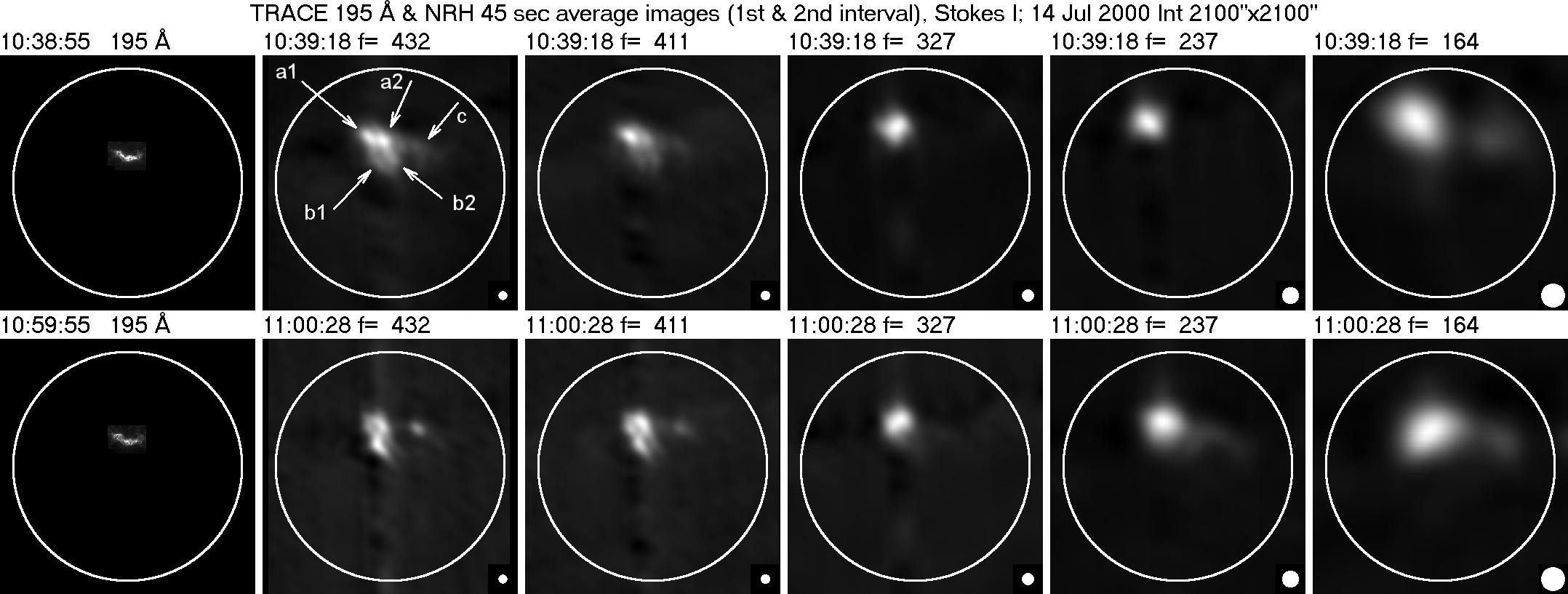}
\end{center}
\caption{Average NRH images in total intensity (Stokes I)  during the first (top) and the second (bottom) interval, together with the corresponding TRACE images in the 195\,\AA\ band (left column). For the NRH images we used a circular clean beam in order to enhance some features. All images are normalized to their peak intensity}
\label{NRH_2D}
\end{figure*}

The apparent velocity on the plane of the sky ranged from 200\kms\ at \frqcM\ to 300\kms\ at 236.6\,MHz, thus the distance of the \frqcM\ source from the 236.6\,MHz\ source increased with time. On the other hand, the relative frequency drift, measured from the delay of the bulk of the emission among the four highest NRH frequencies was $-3.1\,\times10^{-3}$\,s${-1}$; for a hydrostatic isothermal corona at $1.4\times10^6$\,K and a height of 100\,Mm above the photosphere, this drift corresponds to a radial speed of 500\,\kms; to make this value compatible with the apparent speed requires an inclination of $\sim30$\degr\ of the radial direction with respect to the sky plane.

From the above description we conclude that the fiber bursts first appeared in a moving type IV source, which developed after the primary energy release. We note that this moving type IV is not the same as the one described by \cite{2001A&A...373.1073K}, which occurred earlier in the North-West quadrant of the disk. The appearance of the fibers was probably associated with the rapid expansion of the flare ribbons and the formation of post-flare loops in the eastern part of the flaring region, while an association with the eastern footpoint of the CME flux rope is possible.

\subsection{Structure of the fiber-emitting sources}

Average NRH images during the two selected intervals of spike activity are shown in Fig. \ref{NRH_2D}, together with the corresponding TRACE images at 195\,\AA. We used the full disk EIT images to compute an accurate pointing of the TRACE images, which should be good to a few arc seconds. As for the NRH, the absence of image jitter (see next section) shows that the pointing is stable, whereas absolute pointing offsets due to ionospheric refraction are expected to be small during the summer.

During the first interval the \frqc\ and \frqaM\ emission was dominated by two sources, marked {\it a1} and {\it a2} in the figure. These sources were located at the tip of two parallel, low intensity stripes ({\it b1} and {\it b2} in the figure) extending to the South-West; there was also a weak source, {\it c} in the west. Although there is considerable structural change between the two intervals ({\it c.f.} Movie 1), the same sources are detectable during the second interval. At \frqaM\ sources {\it a1} and {\it a2} merge, the stripes are not detectable, while source {\it c} is hardly visible. The convolution of the high frequency images with the \frqaM\ beam showed that these changes  are real structural changes, rather than spatial resolution effects. 

\begin{table}[h]
\caption{Average brightness temperature and polarization of NRH sources during the two selected intervals}
\label{Table:sources}
\begin{center}
\begin{tabular}{l|rrrr|rrrr}
\hline
Freq&\multicolumn{4}{c}{$T_{\rm b}$, ($10^{8}$\,K)}&\multicolumn{4}{c}{$p$, (\%)}\\
(MHz) &   a1 &  a2 &  b2 &  c  &   a1 &  a2 &  b2 &  c  \\
\hline
    & \multicolumn{8}{c}{Interval 1}       \\
\hline
164.0 &\multicolumn{2}{c}{~~5.3}&   - &   - &\multicolumn{2}{c}{20}&  -  &  -  \\
236.6 &\multicolumn{2}{c}{ 32.5}&   - &   - &\multicolumn{2}{c}{50}&  -  &  -  \\
327.0 &\multicolumn{2}{c}{ 11.0}&   - &   - &\multicolumn{2}{c}{50}&  -  &  -  \\
410.5 &        4.9       & 2.9  & 1.4 & 0.8 &     85     &    64   & 28  &  -  \\
432.0 &        2.3       & 2.4  & 1.0 & 0.6 &     62     &    61   & 40  &  -  \\
\hline
    & \multicolumn{8}{c}{Interval 2}       \\
\hline
164.0 &\multicolumn{2}{c}{~~2.8}&   - &   - &\multicolumn{2}{c}{71}&  -  &  -  \\
236.6 &\multicolumn{2}{c}{~~2.5}&   - &   - &\multicolumn{2}{c}{71}&  -  &  -  \\
327.0 &\multicolumn{2}{c}{~~4.0}&   - &   - &\multicolumn{2}{c}{66}&  -  &  -  \\
410.5 &        2.4       & 3.2  & 1.4 & 0.9 &     20     &    72   & 28  &$-85$\\
432.0 &        1.7       & 2.6  & 2.9 & 1.4 &     20     &    75   & 80  &$-98$\\
\hline
\end{tabular}
\end {center}
\end{table}

The peak observed brightness temperature, $T_b$, and the degree of circular polarization, $p$, of sources {\it a1, a2, b2} and {\it c} are given in Table~\ref{Table:sources}. The radio sources are resolved, however their true size might be smaller than the observed (and hence their $T_b$ higher), due to scattering of the radiation in the corona. In Interval 1 the observed $T_b$ was much higher at at 327.0 and 236.6\,MHz than at other frequencies, reflecting the fact that these frequencies were near the peak of the drifting continuum discussed above (Fig. \ref{FullASG}). 

\begin{figure}[h]
\begin{center}
\includegraphics[width=.45\hsize]{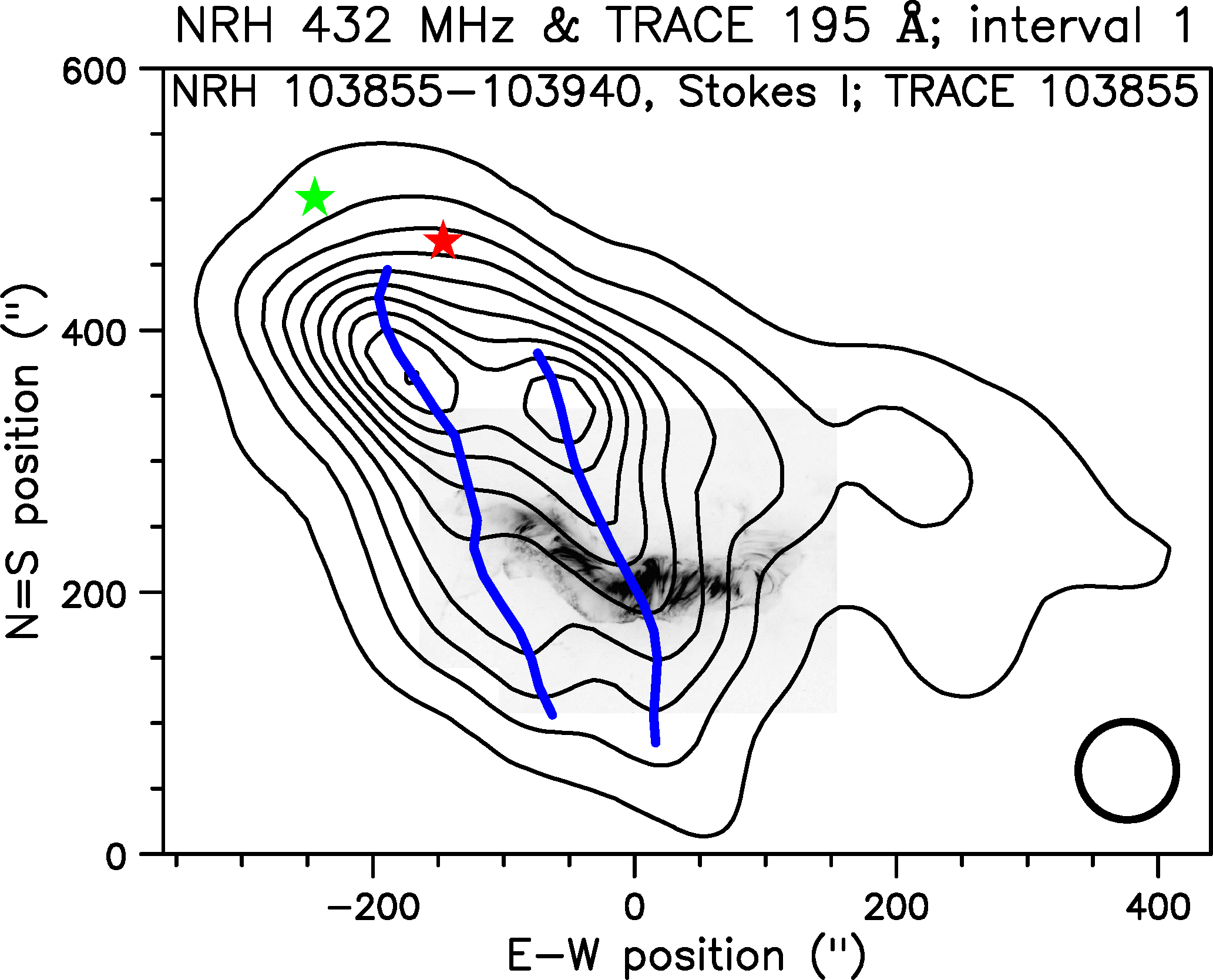}~~~\includegraphics[width=.45\hsize]{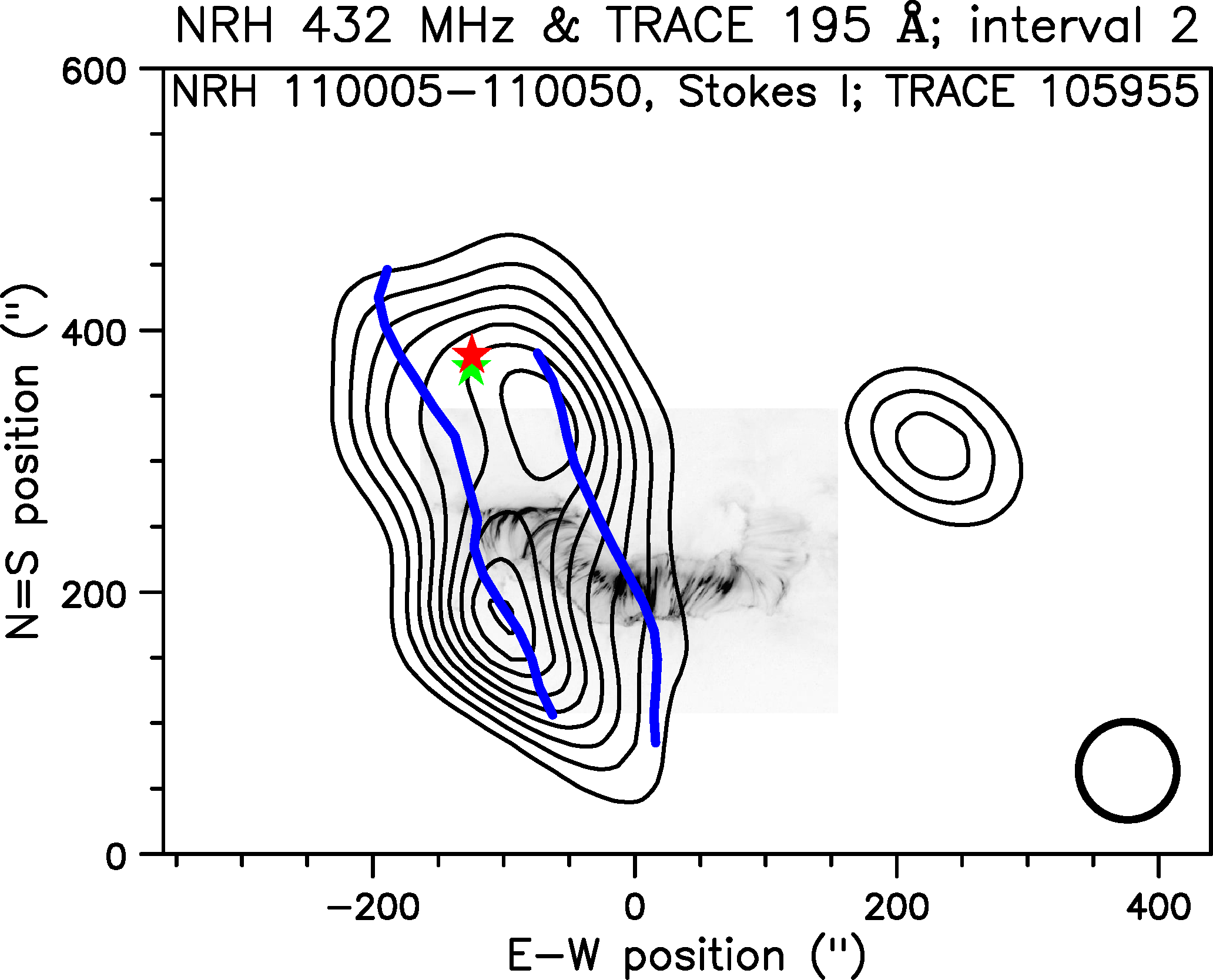}
\end{center}
\caption{Contours of the average NRH images at 432\,MHz during interval 1 (left) and interval 2 (right) superposed on top of the corresponding (negative) 195\,\AA\ TRACE images. The blue lines mark the position of the two stripes during the first interval. The red and green stars mark the position of the \frqa\ and 237.0\,MHz peaks respectively. The NRH beam is plotted in the lower right corner}
\label{contr_avr}
\end{figure}

\begin{figure*}
\begin{center}
\includegraphics[width=.75\hsize]{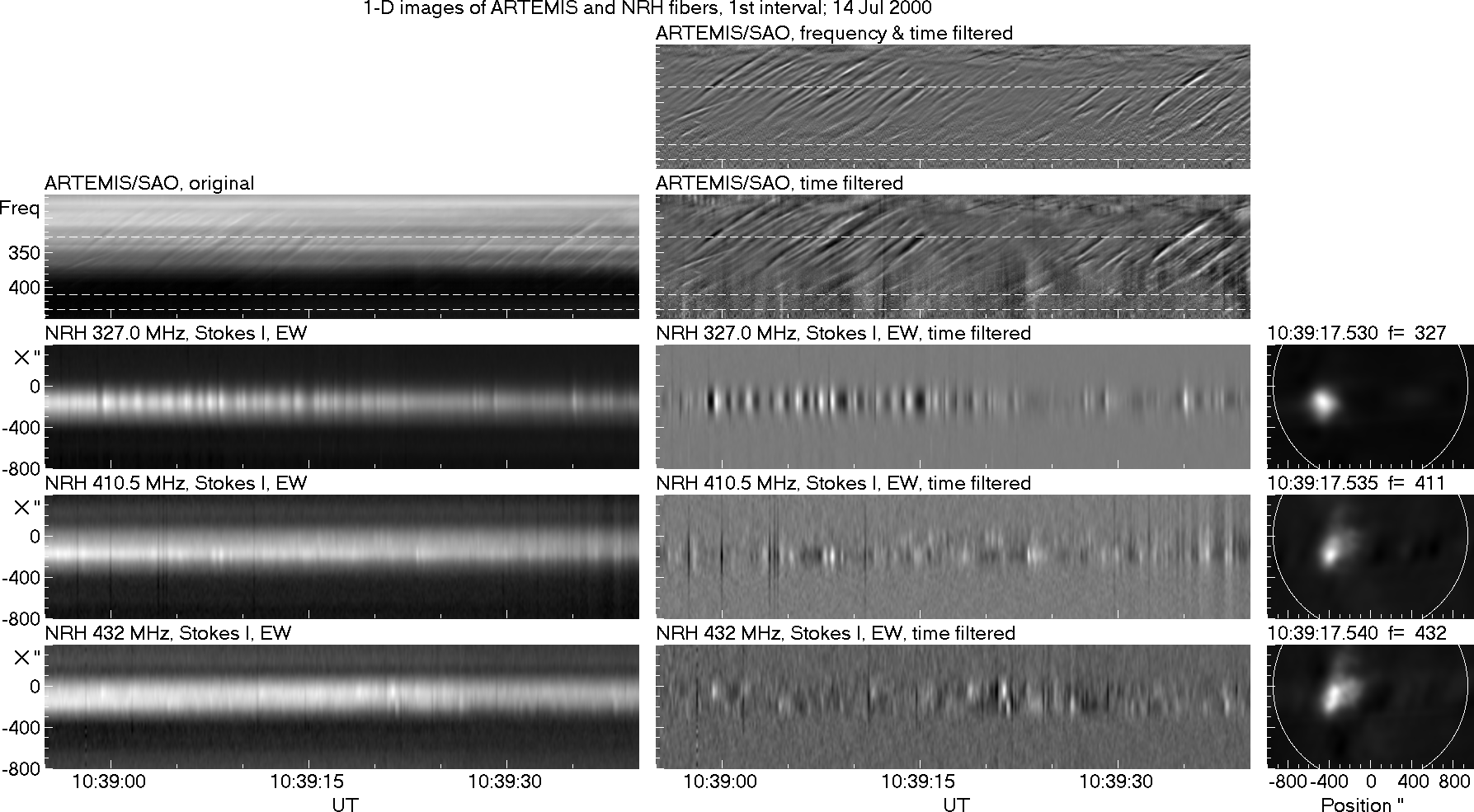}
\end{center}
\caption{Left column: Original SAO dynamic spectrum and 1D Nan\c cay EW images (scans) as a function of time for Interval 1. Middle: The corresponding filtered images; in the top row the dynamic spectrum has been filtered in both time and frequency, in the others in time only. Right: Average NRH images with the same orientation as the 1D images (west limb up). Dashed horizontal lines in the spectra mark the NRH frequencies.}
\label{1D_first}
\end{figure*}

All bright sources were strongly polarized, 50\% or more, at all frequencies with the exception of 164\,MHz in Interval 1; it thus appears that the emission at this frequency, which is well outside the drifting continuum, is part of a different structure. With the exception of source {\it  c} during Interval 2, the polarization is in the right circular sense which, for o-mode emission, corresponds to  magnetic polarity south of the neutral line. 

The relative positions of the metric and EUV emissions are shown better in Fig. \ref{contr_avr}. Here the EUV loop arcade serves in marking the neutral line of the magnetic field. In the same figure we have plotted the positions of the stripes, measured on the \frqcM\ image during the first interval (blue lines). The distance between the stripes was about 100\arcsec, which is about half of the EW extent of the loop arcade.

The two stripes appear to define segments of large scale loops, with projected length of about 300\,Mm, located above the post-flare loops.  It is thus certain that the continuum source and the fibers were not associated to the 195\,\AA\ post-flare loops, but to a much larger magnetic structure. The comparison of the orientation of the large-scale loops with that of the post-flare loops, shows that the former are inclined eastwards with respect to the latter; the same conclusion is drawn from the comparison with the model loops computed by \cite{2001ApJ...551L.115Y}.

During the first interval, sources {\it a1} and {\it a2} together extended in longitude as much as the 195\,\AA\ loop arcade, shifted with respect to that on the plane of the sky by $\Delta x=-60$\arcsec, $\Delta y= 100$\arcsec\ (total $\sim120$\arcsec). This shift is not compatible with a radial displacement of sources {\it a1} and {\it a2} with respect to the flare and strengthens the view expressed in the previous paragraph the magnetic structure is inclined eastwards.

Going to Interval 2, we note that source {\it a1} is weak and {\it b2} even weaker than during interval 1 and the emission is dominated by sources {\it b1} and {\it a2}, whose projected positions span the neutral line. At first sight, this might suggest that they are located near the legs of a magnetic loop; however, as shown in Fig. \ref{contr_avr}, {\it b1} is part of the east stripe and {\it a2} is part of the west stripe. This is corroborated by the fact that they are both polarized in the same sense, which indicates that they are both located in regions of the same magnetic polarity.

Source {\it c} is well visible during Interval 2 and is the only one with left-hand circular polarization. It is located at some distance from the west end of the flaring loop arcade, at the extension of the neutral line. We could not find any association of this source with magnetic or EUV features. During the first interval, source {\it c} is not well resolved, it appears to connect with source {\it a2} (see frame at 10:39:45, Movie 1) and its polarization was too low to be measured. We note that a very strong source existed earlier in the event, between 10:29 and 10:33 UT, at the same location as source {\it c} (see Movie 1), but its relation with source {\it c} is not clear.

\subsection{Imaging of individual fibers}\label{Sect3.4}
	The most convenient way to visualize fine structures present in the NRH images is to compute 1D images (scans) and compare them with the dynamic spectrum, as in \cite{Bouratzis2016}. For this purpose the SAO data were integrated over 0.06\,s, in order to approach the NRH time resolution of 0.125\,s; they were also subjected to a 5\,s wide high pass Gaussian filter in the time domain and a 20\,MHz wide high pass Gaussian frequency filter, in order to suppress large scale structures \citep{Bouratzis2016}. Furthermore, the filtered dynamic spectra were normalized by dividing the intensity of each frequency channel by its rms value. The 1D NRH scans were subjected to the same time filter but, obviously, could not be subjected to frequency filtering. Shorter time filters gave similar results down to filter widths of $\sim1$\,s, which is close to the duration of the absorption-emission fiber pair.

Fig. \ref{1D_first} shows the original and time-filtered 1D NRH intensity as a function of time for the first interval, with the 1D images computed in the EW direction; in the same figure we give the SAO dynamic spectrum, (original, filtered in frequency and time, and filtered in time only), as well as the averaged NRH 2D images for reference. We first note that, although fibers are hard to see in the original spectrum, they are readily visible in the filtered spectra. A similar improvement is noticed when comparing the time-filtered NRH 1D images to the original ones. We further note that, in the dynamic spectra filtered in time only, broad band emissions do not significantly affect the fibers; we expect that the same thing is true for the 1D filtered images. 

\begin{figure}
\begin{center}
\includegraphics[width=\hsize]{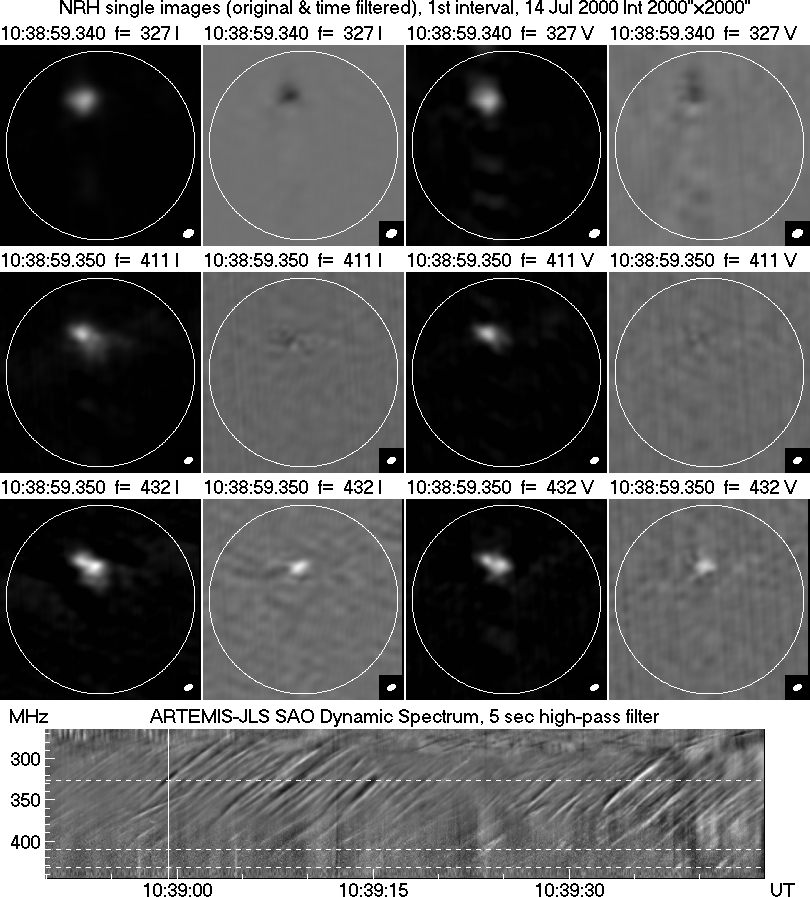}
\end{center}
\caption{A frame from Movie 2, which shows the full sequence of 125\,ms NRH images during Interval 1 at 327.0, (first row from the top), 410.5 (second row) and 432.0\,MHz (third row). Each row shoes the original and time-filtered images in Stokes I and V; right-hand circular polarization is white. 
A linear gray-scale color table was used for the displays, where black corresponds to the minimum and white to the maximum intensity during the entire interval of each image set.
At the bottom we show the time-filtered SAO spectrum, extended by 5\,s on either side; the vertical line marks the time of the images and the dashed horizontal lines the NRH frequencies.}
\label{Single_M1}
\end{figure}

In spite of the lower time resolution of the NRH,  practically all fiber bursts in the SAO spectra are detectable in the NRH 1D scans. They  appear as short duration enhancements above the slowly varying background, rather than as discrete sources, which was the case with spikes \citep[{\it c.f.} Fig. 16 of][]{Bouratzis2016}. In this interval, fibers are best visible at 327.0\,MHz, being rather weak but still discernible at the higher frequencies. None of the 327.0\,MHz fibers crossed the higher NRH frequencies, there are, however, a few fibers crossing both the 410.5 and 432.0\,MHz levels.

The shape and position of fibers are best compared with the background source in 2D time-filtered images, which were computed by applying the high-pass time filter to each pixel of the original image set. Fig. \ref{Single_M1} gives a frame of the associated Movie 2, which shows all 2D NRH images in Interval 1, original and time-filtered, in Stokes I and V. In the frame shown in the figure we have a fiber in absorption at \frqaM, no fiber at \frqbM\ and a fiber in emission at \frqcM. Despite the short integration of the individual NRH images, their quality is good, with a noise level of about $3\times10^6$\,K, both in the original and the time-filtered images; we have only a few bad images, which appear as vertical dark streaks in the 1D images of Fig. \ref{1D_first}.

\begin{figure}
\begin{center}
\includegraphics[width=\hsize]{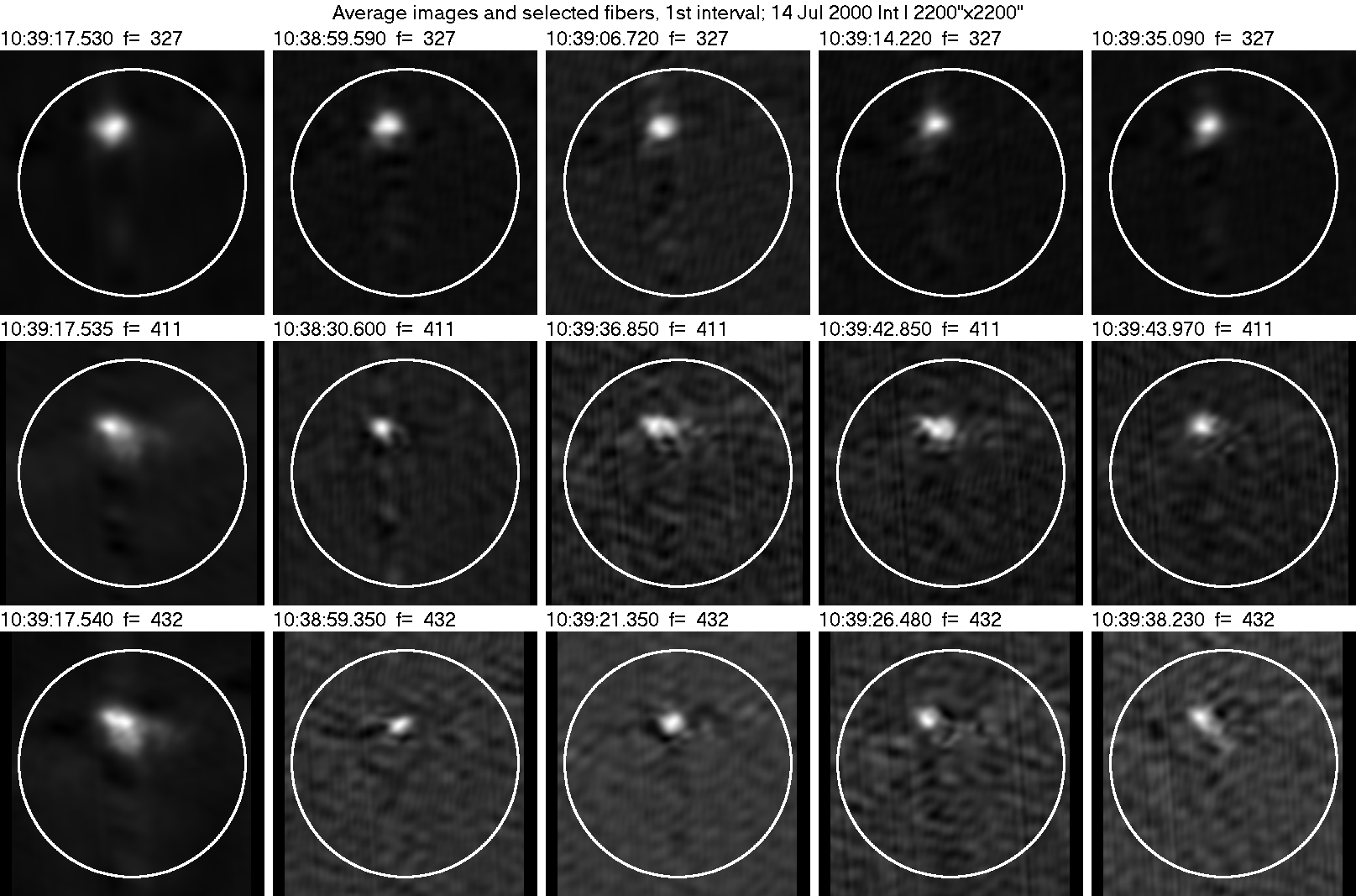}
\end{center}
\caption{A selection of time-filtered NRH images in Stokes I during Interval 1 at \frqa\ (top row), \frqb\ (middle) and \frqcM\ (bottom). The average image at each frequency is shown on the left column for reference.}
\label{Selected1}
\end{figure}

Fig. \ref{Selected1} shows a selection of fiber images from Interval 1 at the three NRH frequencies, together with 45-sec averaged images (left column). Only total intensity images are given since the sources are uniformly polarized, but both I and V images are included in Movie 2. At \frqcM\ (bottom row in Fig. \ref{Selected1}), the emission comes from the western source ({\it a2}) at 10:38:59.350 and 10:39:21.350 UT, from the eastern source ({\it a1}) at 10:39:26.480 UT and from both sources {\it a1} and {\it b2} at 10:39:38.230 UT. These differences in position are also visible in the 1D scans of Fig. \ref{1D_first}. Similarly, at \frqbM, source {\it a1} alone is implicated at 10:38:30.600 and 10:39:43.970 UT, both {\it a1} and {\it a2} at 10:39:36.850 UT and mostly source {\it a2} at 10:39:42.850 UT. At \frqaM\ (top row), the size and position of the fiber emission is very close to those of the single continuum source.

\begin{figure*}
\begin{center}
\includegraphics[width=.75\hsize]{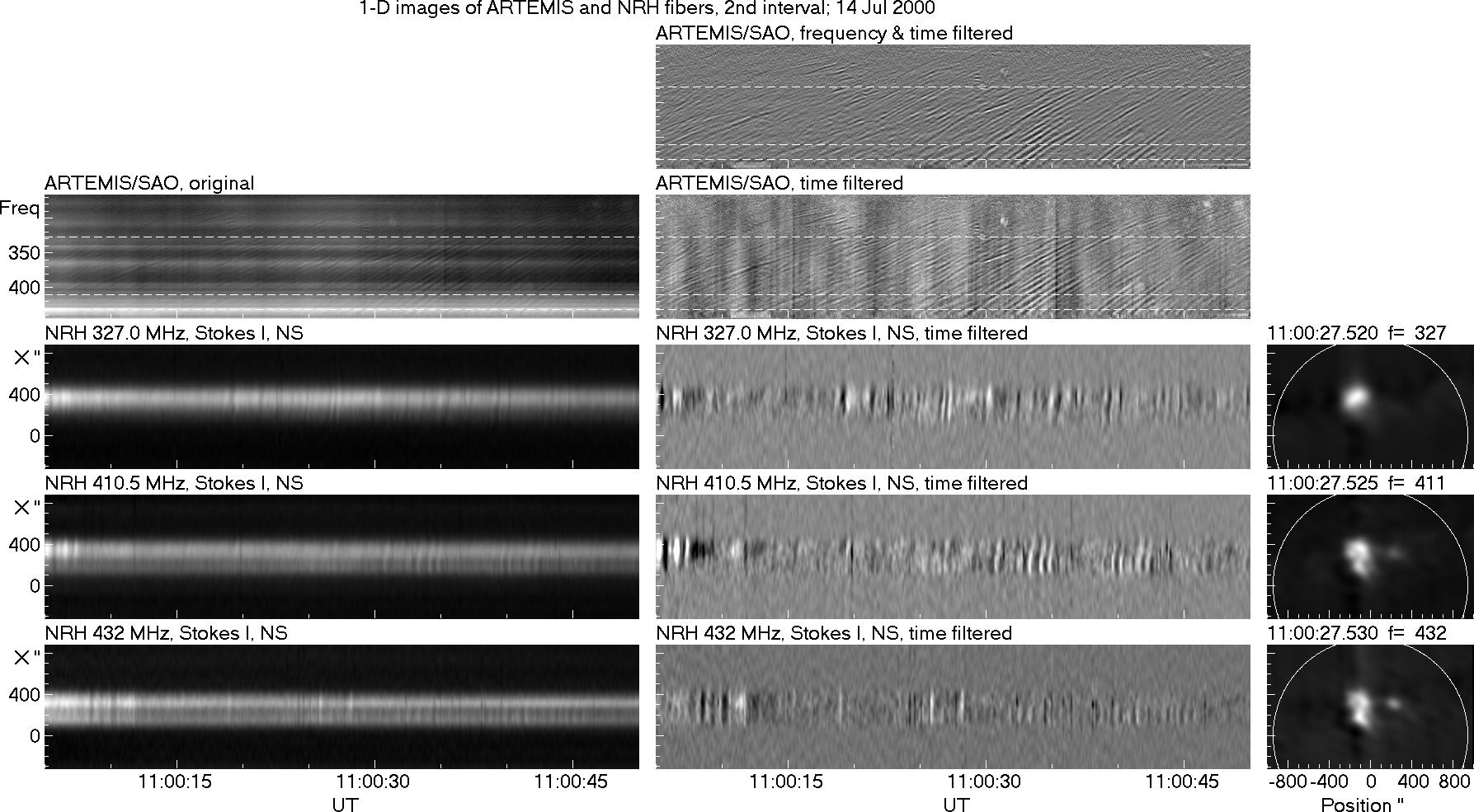}
\end{center}
\caption{Same as Fig. \ref{1D_first} for the second interval, with the 1D scans in the NS direction. The arrow marks fibers exhibiting apparent motion. The average images at right are oriented with the north up.}
\label{1D_second}
\end{figure*}

\begin{figure}
\begin{center}
\includegraphics[width=\hsize]{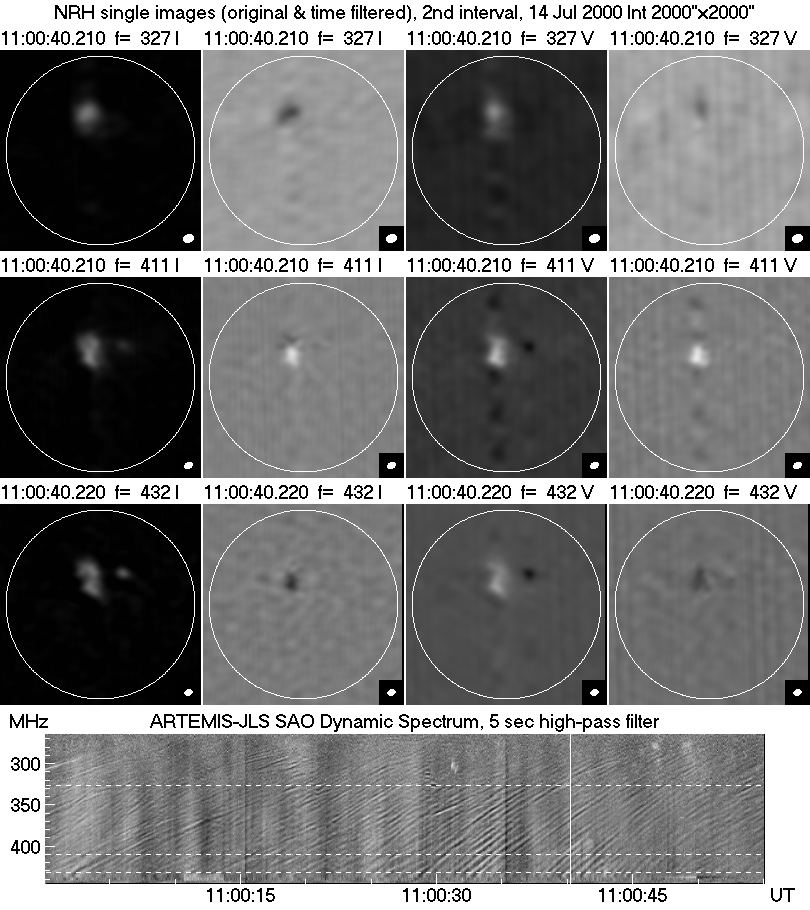}
\end{center}
\caption{Same as Fig. \ref{Single_M1}, showing a frame from Movie 3 (Interval 2).}
\label{Single_M2}
\end{figure}

One-dimensional NRH scans for the second interval, this time along the NS direction, are given in Fig. \ref{1D_second}. Here fibers are also well visible both in the time filtered dynamic spectrum and the time filtered 1D scans; they are stronger at the high NRH frequencies than at \frqaM, which is apparently a consequence of the fact that the continuum source moved to lower heights, as mentioned previously (Sect. \ref{ContEvol}). Many fibers cross both the 432.0 and 410.5\,MHz frequency levels, but hardly any cross all three. The full evolution is given in Movie 3, a frame of which is shown in Fig. \ref{Single_M2}, with fibers in absorption at \frqaM\ and \frqcM\ and in emission at \frqbM. 

\begin{figure}[h]
\begin{center}
\includegraphics[width=\hsize]{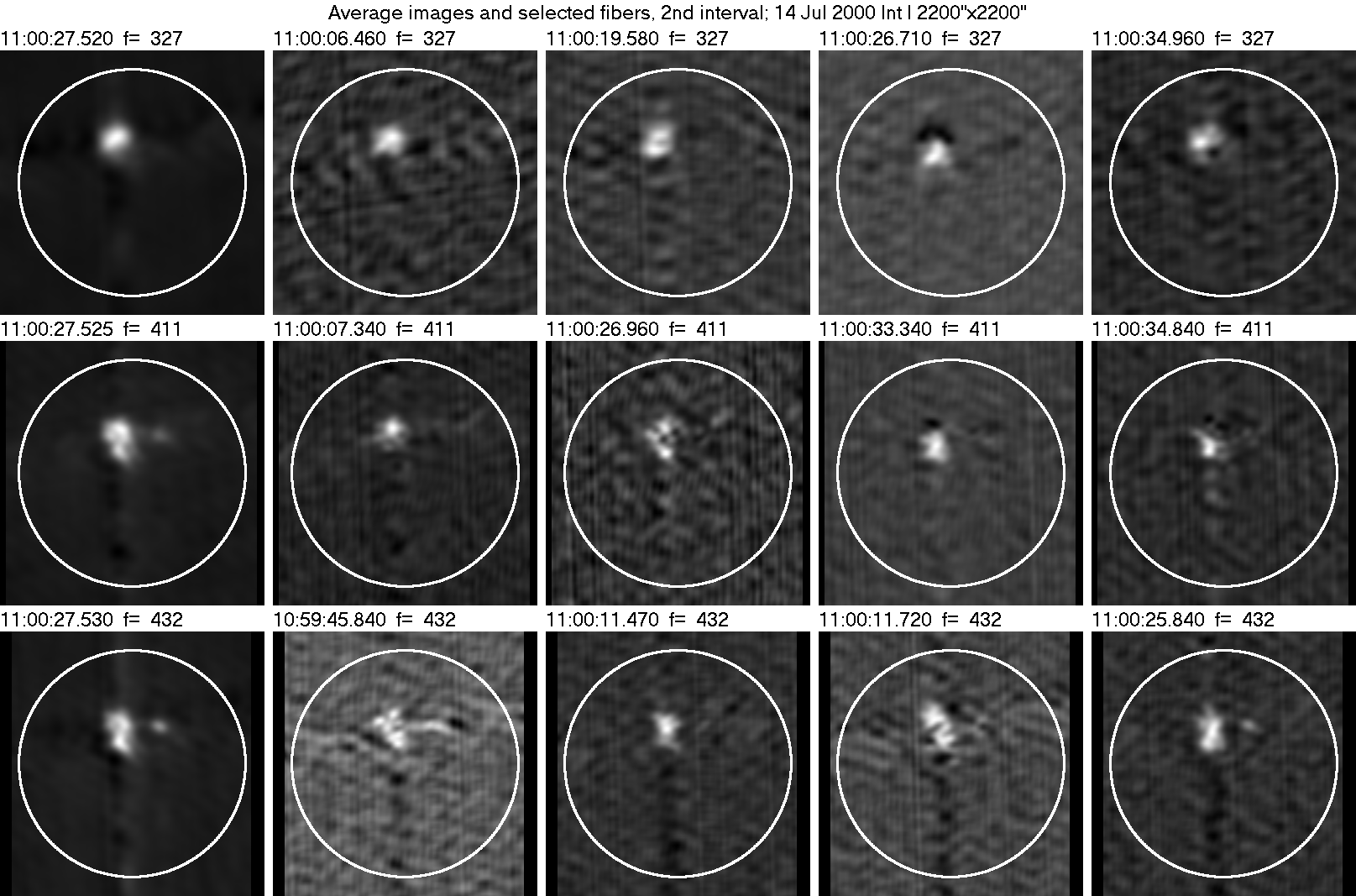}
\end{center}
\caption{A selection of time-filtered NRH images in Stokes I during Interval 2. The average image at each frequency is shown on the left column for reference.}
\label{Selected2}
\end{figure}

It is quite obvious in the 1D scans of Fig. \ref{1D_second} that in most cases the fiber emission extended over both sources {\it a1-a2} and {\it b1-b2}. This is better shown in the selected images of Fig. \ref{Selected2}, where we have cases of fibers from sources {\it a2, a2} and {\it b1}, as well as cases of emission from a single continuum source in between. 

We note that source {\it c} showed occasional fiber activity, as in the \frqcM\ image at 11:00:25.840 UT (Fig. \ref{Selected2}); moreover, in one case (10:59:45.840 UT at \frqcM, also in Fig. \ref{Selected2}), there appears to be a connection between source {\it c} and the others. Taking this into consideration, as well as its polarization, source {\it c} might represent the other foot point of the large scale loops. Apart from this occasional activity, source {\it c} was very stable both in intensity and in position and did not show up in most of the time-filtered 2D images (Movie 3); this shows, among others, that our data are free of jitter due to the terrestrial atmosphere/ionosphere, at least for times scales shorter than the 5-sec width of the applied filter. 

The time sequences of 1D NRH images are convenient for checking fluctuations in circular polarization. Time variations in Stokes parameter V follow closely those in total intensity. Fluctuations in the degree of circular polarization are considerably weaker than fluctuations in I or V and, in general, in phase with them. Thus the degree of polarization of fibers in absorption is a few percent smaller than that of fibers in emission, a difference which we do not consider significant.

\subsection{Apparent motions in individual fibers}\label{AppMot}
In many cases, fibers in the 1D intensity-time displays (Figs. \ref{1D_first} and \ref{1D_second}) have the form of vertical streaks, which implies that the emission appears simultaneously over the entire structure. There are, however, cases where the fiber signatures are inclined, indicating apparent motions of individual features on the sky plane. One example is near 11:00:32 UT in the NS cuts of Fig. \ref{1D_second}; similar inclinations were detected in the EW cuts of the same fibers (not shown here.) The inclinations suggest apparent motions from West to East and from South to North, {\it i.e.} from from the lower to the upper part of the loops associated with the two stripes. The apparent velocity, measured by 2d autocorrelation, is supra-luminal, ranging from 4c to 10c.  

\begin{figure}
\begin{center}
\includegraphics[width=\hsize]{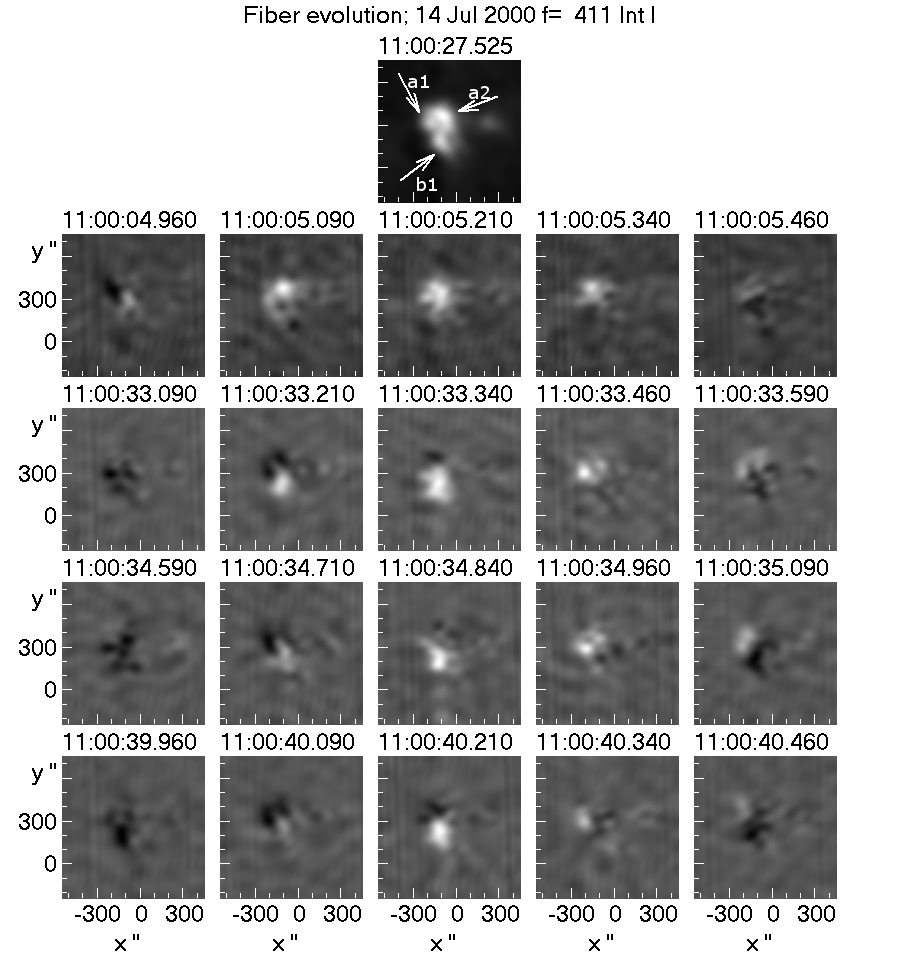}
\end{center}
\caption{Time sequences of filtered NRH 2D images, showing the evolution of four fibers at \frqbM\ during the second interval. The time step is 125\,ms. At the top row we give the corresponding average image for reference.}
\label{motions}
\end{figure}

\begin{figure}
\begin{center}
\includegraphics[width=\hsize]{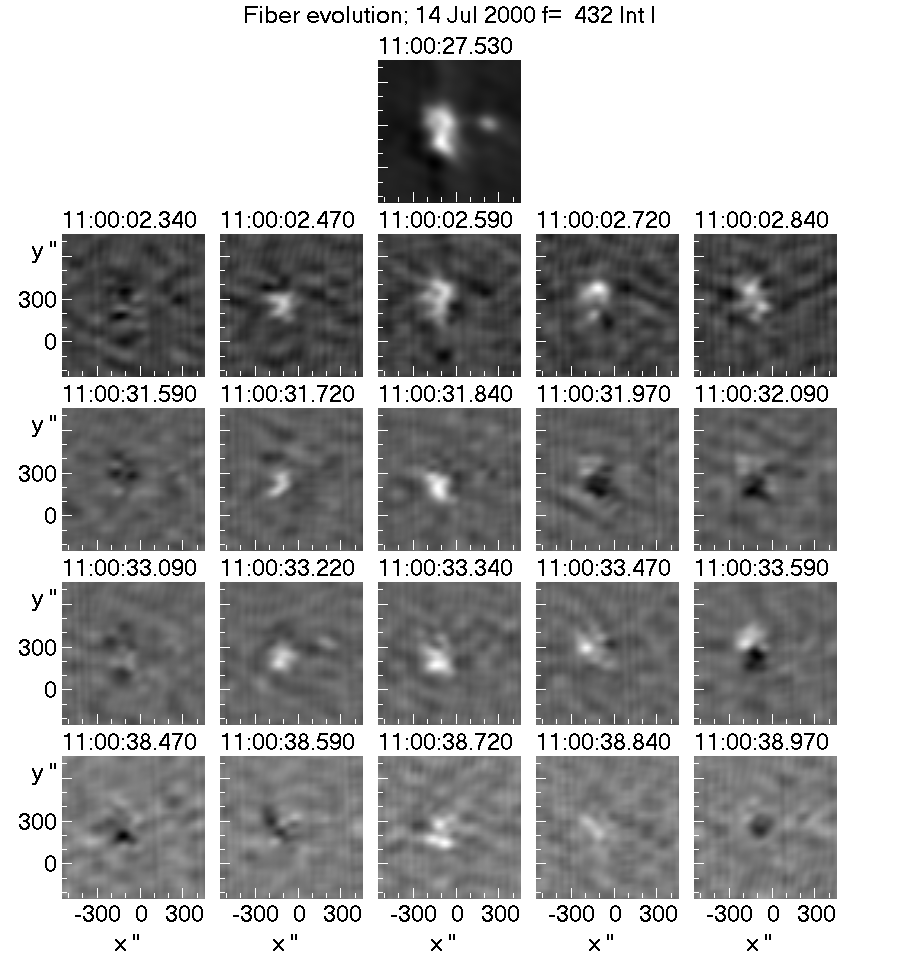}
\end{center}
\caption{Same as Fig. \ref{motions} for  \frqcM.}
\label{motions432}
\end{figure}

In Fig. \ref{motions} we present 5-image (500\,ms) long sequences of filtered 2D images for four fibers at \frqcM. The one around 11:00:05 UT (first row below the average image) was practically stationary, while the other three showed clear apparent motions from South-West to North-East, in conformity with the results from the 1D image analysis. We first note that in some of the images shown in this figure, as well as in many others, fibers in absorption coexist with fibers in emission. Thus in the first image of the bottom row we have absorption at the location of source {\it b1}. In the next frame this absorption shrinks and we have new absorption at the location of sources {\it a1}and {\it a2} and one image later {\it b1} appears in emission. In the fourth image the emission at {\it b1} disappears and we have emission from source {\it a1}, while in the last image we have new absorption at the location of source {\it b1}. Similar is the situation with the time sequences starting at 11:00:34.59 and at 11:00:33.09 UT, except that in these we also have emission from source {\it a2} in the fourth frame.

We conclude that what in the dynamic spectrum appears as a single fiber is actually a complex of multiple wave trains that cross the plasma level at slightly different times and positions, giving the impression of an apparent motion which may exhibit supra-luminal velocity. Fig. \ref{motions} indicates that the duration of each train is comparable to the time difference between successive trains, which explains why such trains do not appear as separate fibers in the dynamic spectrum. It is, however, rather peculiar that the pattern of trains repeats itself several times, as indicated by the similarity of the time evolution of the three fibers of Fig. \ref{motions}.

\subsection{Fiber images at two frequencies}
The best observed cases of fibers crossing two NRH frequencies were during the second interval and for the closely spaced NRH frequencies of \frqcM\ and \frqbM. In Fig. \ref{motions432} we give 5-image sequences at \frqcM\ of the same fibers that are shown in Fig. \ref{motions}. The time delay between the two frequencies is 2.75\,s for the first fiber and 1.5\,s for the other three, corresponding to frequency drifts of $-7.8$ and $-14.3$\,MHz\,s$^{-1}$ respectively.

Although the images of the same fiber at the two frequencies are similar, they are not identical. Differences could be attributed to the fact that the actual time delay is not an integer multiple of the image cadence, so that evolution effects including the apparent motions discussed in the previous section play a role in the form of the radio sources. 
For the same reasons, the (small) differences in the position of individual fibers between \frqc\ and \frqbM\ cannot be accurately measured. 

\subsection{Exciter speed and frequency scale length}
Although, as mentioned previously, we could not identify any fibers crossing all three NRH frequencies, from 11:00:29 to 11:00:45 (second interval) we had a pattern of fibers that varied smoothly between \frqcM\ and \frqaM\ (Fig. \ref{Single_M2}). Using the method of cross correlation with a sliding spectral window described in Section 2.2.2 of paper I, we computed the average track of this group of fibers on the dynamic spectrum, and from that we obtained time delays of 1.74 and 8.52\,s for \frqbM\ and \frqaM, with respect to \frqcM. 

We also measured the position shift, considering that the average 2D images, taking into account the time delays, represent well the position of the fiber sources. By convolving these images with the \frqaM\ beam and measuring the source positions through a least square fit of sources {\it a1} and {\it a2} together with a single Gaussian, we obtained shifts of 18\arcsec\ and 58\arcsec\ for \frqbM\ and \frqaM, with respect to \frqcM. From these measurements, we computed the speed of the exciter on the plane of the sky and found a value of 5\,Mm\,s$^{-1}$, in good agreement with the results of paper I (Fig. 17, left histogram), computed from the frequency drift rate. 

From the position shifts and the corresponding frequencies we also computed the frequency scale length (twice the density scale length) along the loop, projected on the plane of the sky, $\ell_f$. For an isothermal loop, the plasma frequency varies with distance, $x$, along the loop as:
\be
f=f_0 \exp{\left(-\frac{x-x_0}{\ell_f}\right)}
\ee 
We obtained a value of $\ell_f=146$\,Mm. On the other hand, the value of the frequency scale length along the loop, $h_f=H_f/\cos\theta$, where $H_f$ is the scale in the radial direction and $\theta$ the angle between the loop segment and the vertical, can be computed by fitting the average fiber track to the expression (11) of paper I. This expression was derived for a simple model of the fiber tracks on the dynamic spectrum; it is independent of coronal parameters, but presumes whistler origin of the emission. We obtained $h_f=183$\,Mm.

The quantities $\ell_f$ and $h_f$ are related through the simple expression:
\be
\ell_f=h_f \cos \beta
\ee
where $\beta$ is the angle between the line of sight and the loop segment, and the above measurements give $\beta=37$\degr. For a simple geometrical model of a vertical semicircular loop, we have
\be
\beta=90\degr-\alpha-\theta
\ee
where $\alpha$ is the angle between the line of site and the radial direction and $\theta$ was defined above. Taking $\alpha\sim16$\degr\ (equal to the flare latitude), we get $\theta\sim37$\degr, which places the fiber-emitting sources {\it a1} and {\it a2} above the middle of the leg of the large-scale loop.

Thus the 2D images and dynamic spectrum give consistent results both for the exciter velocity and for the frequency scale length. Moreover, we have one more element in favor of whistler origin of fibers, since the analysis of spectral data that we used to measure the frequency scale was based on that assumption.

\subsection{Fibers in emission and absorption}
According to the theory of whistler origin ({\it cf.} Sect. \ref{intro}), the fiber emission appears at 
$f_{pe} + f_w$ and the absorption at $f_{pe}$. 
Thus, at a fixed frequency, the absorption will precede the emission by
\be
\Delta t=-f_w \left( \frac{df}{dt}\right)^{-1}
\ee
where $df/dt$ is the frequency drift rate. The whistler frequency is equal to the frequency difference of the absorption and emission ridges, hence $\Delta t$ can be computed from the dynamic spectrum. It amounts to 350\,ms for Interval 1 and 250\,ms for Interval 2.

\begin{figure}[h]
\begin{center}
\includegraphics[width=\hsize]{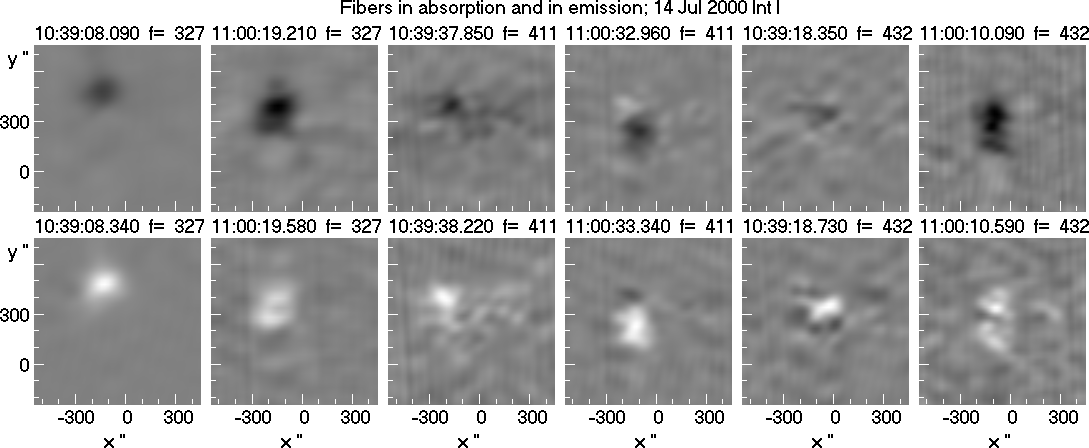}
\end{center}
\caption{Time-filtered images of fibers in absorption (top row) and in emission (bottom). One case is shown for each interval and each NRH frequency. Each pair of fiber images is normalized to the same minimum/maximum values.}
\label{AbEm}
\end{figure}

Fig. \ref{AbEm} shows some examples of fiber pairs in absorption and in emission; two pairs are presented for each frequency, all from Interval 1. The time difference is from 250\,ms to  500\,ms (2-4 NRH images), in conformity with the prediction of the previous paragraph. We note that the shape and size of the absorption and emission of each fiber pair are vary similar, which verifies the hypothesis that they are manifestations of the same wave train. The difference of their intensities from the background are similar, although exact measurements are uncertain, as the intensity values might be affected by the filtering process. Their position is also very close, but it cannot be measured with certainty due to the apparent motions that some fibers exhibit (Sect. \ref{AppMot}).  

\section{Summary, discussion and conclusions}\label{conclusions}
The combined spectral/imaging observations with ARTEMIS-JLS/SAO and the NRH revealed a number of important aspects of fiber bursts observed during the July 14, 2000 event, which we summarize and discuss in this section.

We found that the fibers modulate the intensity of the radio emission by about 10\%, below the 30\% to 40\% range reported by \cite{2017ApJ...848...77W} at higher frequencies, thus images averaged over several seconds are representative of the background emission. We used such images to study the evolution of the continuum and its association with the fibers.

The first fibers were detected in a moving type IV source, which in the dynamic spectrum manifested itself as a broad band drifting continuum that started after the GOES and HXR  peaks. This emission occurred North-East of the flare and was probably associated to the rapid eastward expansion of the flare and the associated loop arcade, that occurred at about the same time; that was after the main energy release and the filament activation, which occurred in the western part of the active region. The fibers appeared near the peak intensity of the drifting continuum and close to the time that the radio sources started an upward rectilinear motion with a projected velocity of 200-300\,km\,s$^{-1}$ that lasted for 4-6 min. After this upward motion, the continuum source moved in an irregular way to the South-East, approaching the eastern part of the post-flare loop arcade. This downward motion could be the result of the decrease with time of the ambient electron density in the associated magnetic structures, so that the plasma level corresponding to the observing frequencies moves to lower heights.

We selected for further study two 45-sec intervals that were not contaminated by pulsations and other broad-band structures. Average images, representing the background continuum, revealed two parallel stripes, {\it b1} (east) and {\it b2} (west), apparently segments of large-scale loops, much larger than the EUV post flare loops, and  inclined eastward with respect to them. This indicates that the plasma that produced the type IV and the fiber emission was confined in such loops and was not associated with the post-flare loops seen in the EUV. We note that
\cite{2005A&A...435.1137A}
found that the lines of force of their extrapolated magnetic field were higher than the soft X-ray flare loops. A similar conclusion is drawn from Fig. 7a of \cite{2007SoPh..245..327R}. On the other hand, \cite{2017ApJ...848...77W} reported that the fiber sources  around 1400\,MHz were near and above one footpoint of the flare loops; this might be due to the fact that their observing frequency was considerably higher than that of the NRH, hence plasma emission should originate from a higher density region, located lower in the fiber loop.

The fact that all sources implicated in the fiber emission were polarized in the same sense supports the geometry suggested above. The observed polarization was right handed, in the sense of the ordinary mode south of the neutral line. The mere presence of fiber bursts indicates plasma emission and the strong polarization points to emission at the fundamental, which is expected to be highly polarized in the ordinary sense because the extraordinary mode is evanescent near this frequency.

The whistlers are excited as a result of a loss-cone instability \citep{Kuijpers1972}, the development of which depends on the value of the loss-cone aperture angle. In the case of an asymmetric loop, it is possible that the instability will  develop in one leg of the loop as, according to \cite{2005A&A...435.1137A}, conditions are more favorable at the weak field footpoint. This probably explains why we do not see emission from both legs of the large scale loops.

A possible explanation of the large size of the stripes, compared to that of the EUV loops, is that they are portions of large scale loops encompassing both the CME-associated flux rope and the EUV flare loops \citep[see, e.g. Fig.1 of ][]{2005ApJ...630.1133R}. Their location near the east end of the flux rope may explain their large inclination, as the flux rope expands more rapidly in the middle, pushing these loops eastwards.

The projected length of the stripes was about 300\,Mm. Assuming a semi-circular loop, a hydrostatic density model with a temperature of $1.5\times10^6$\,K and a base density four times that of the Newkirk model (Sect. 5.4 of paper I), we obtained from the fiber tracks in the dynamic spectrum a radius very close to the measured projected length.

In order to facilitate the comparison of the fiber signatures in the images and the dynamic spectrum at full NRH time resolution, we applied a 5\,s wide Gaussian high pass filter in the time domain to suppress slowly varying emissions; for the dynamic spectra, we also applied a 20\,MHz wide Gaussian high pass filter in the frequency domain to suppress broad band features. Furthermore, for the three NRH frequencies that were inside the SAO band (\frqa, \frqb\ and \frqcM), we computed both 2D images from the original visibilities and 1D images by integration the 2D images in the EW and NS directions. This processing made the fibers very prominent both in the dynamic spectra and in the sequence of images. 

Practically all fibers visible in the SAO dynamic spectra are detectable in the NRH 1D scans and 2D images. We found that some fibers originated in stripe {\it b1} and the associated source {\it a1}, some in {\it b2} and source {\it a2}, and some in between, indicating interactions between the associated loops. We found very small changes of the degree of circular polarization associated with fibers.

A close examination of the 2D time-filtered images revealed cases of multiple fiber emissions appearing at slightly different positions and times. The time differences are comparable to the duration of the emission, thus they appear as single fibers in the dynamic spectrum. The consecutive appearance of such emissions gives the impression of apparent motion with supra-luminal velocities.

Although we found no fibers crossing all  three NRH frequencies that were inside the SAO frequency range, we had a number of fibers crossing the closely spaced frequencies of \frqc\ and \frqbM. Their images were very similar, but we could not reliably measure position differences. Position shifts and the corresponding time delays between \frqa\ and \frqcM\ were measured for a group of fibers that varied smoothly between the two frequencies; our measurements led to estimates of the exciter speed and the frequency scale length, with spectral and imaging data giving consistent results, which supports the hypothesis of whistler origin of the fibers. We also found that the upper sources {\it a1} and {\it a2} were located above the middle of the leg of the large scale loop.

Finally,  we examined fibers appearing in absorption. We found that the shape and size of the absorption and emission fibers are vary similar, which verifies the hypothesis that they are manifestations of the same wave train.

This study confirms the importance of simultaneous spectral and imaging observations of the metric radio emissions. Our analysis revealed the geometry of the continuum and the fiber sources, made possible the direct measurement of exciter speed and frequency scale length, revealed multiple wave trains unresolved in the dynamic spectrum, and provided concrete evidence that fibers in emission and absorption are indeed produced by the same wave train. Moreover, we found that the whistler hypothesis we adopted works well with the imaging results, confirming our conclusion in paper I, based on spectral data. Future work could include non-linear magnetic field extrapolations, as well as the analysis of a larger sample of events to verify the conclusions reached in this work.

\begin{acknowledgements} 
We wish to thank our colleagues from the Observatoire de Paris-Meudon and the NRH staff for providing the original
visibility data. The authors are grateful to the other members of the ARTEMIS group, C. Caroubalos, P. Preka-Papadema, X. Moussas, A. Kontogeorgos and P. Tsitsipis, for their support work.
We also wish to thank A. Nindos and S. Patsourakos for useful discussions and comments. Data from TRACE, GOES, the RSTN network, MTI/HXRS and HXT were obtained from the respective data bases; we are grateful to all that contributed to the operation of these instruments and made the data available to the community.
\end{acknowledgements}

{}


\begin{thebibliography}{}

\bibitem[Aurass et~al.(2005)]{2005A&A...435.1137A}
Aurass, H., Rausche, G., Mann, G., \& Hofmann, A.\ 2005, \aap,  435, 1137 

\bibitem[Bouratzis et al.(2015)]{2015SoPh..290..219B} 
Bouratzis, C., Hillaris, A., Alissandrakis, C.~E., et al.\ 2015, \solphys, 290, 219 

\bibitem[{{Bouratzis} {et~al.}(2016){Bouratzis}, {Hillaris}, {Alissandrakis},
  {Preka-Papadema}, {Moussas}, {Caroubalos}, {Tsitsipis}, \&
  {Kontogeorgos}}]{Bouratzis2016}
{Bouratzis}, C., {Hillaris}, A., {Alissandrakis}, C.~E., {et~al.} 2016, \aap,
  586, A29

\bibitem[{{Bouratzis} {et~al.}(2019){Bouratzis}, {Hillaris}, {Alissandrakis},
  {Preka-Papadema}, {Moussas}, {Caroubalos}, {Tsitsipis}, \&
  {Kontogeorgos}}]
{Bouratzis2019}
{Bouratzis}, C., {Hillaris}, A., {Alissandrakis}, C.~E., {et~al.} 2019, \aap,
625, A58 (paper I).

\bibitem[{{Caroubalos} {et~al.}(2001a){Caroubalos}, {Maroulis}, {Patavalis},
  {Bougeret}, {Dumas}, {Perche}, \& \etal}]{Caroubalos01}
{Caroubalos}, C., {Maroulis}, D., {Patavalis}, N., {et~al.} 2001, Experimental
  Astronomy, 11, 23

\bibitem[Caroubalos et al.(2001b)]{2001SoPh..204..165C} 
Caroubalos, C., Alissandrakis, C.~E., Hillaris, A., et al.\ 2001, \solphys, 204, 165 

\bibitem[{{Elgar{\o}y}(1972)}]{Elgaroy73}
{Elgar{\o}y}, {\O}. 1972, in CESRA-3, Committee of European Solar Radio
  Astronomers, ed. J.~{Delannoy} \& F.~{Poumeyrol}, Vol.~3, 174

\bibitem[{{Benz} \& {Mann}(1998)}]{Benz98}
{Benz}, A.~O. \& {Mann}, G. 1998, \aap, 333, 1034

\bibitem[{{Bernold} \& {Treumann}(1983)}]{Bernold83}
{Bernold}, T.~E.~X. \& {Treumann}, R.~A. 1983, \apj, 264, 677

\bibitem[{{Fomichev} \& {Chertok}(1978)}]{Fomichev78}
{Fomichev}, V.~V. \& {Chertok}, I.~M. 1978, Radiofizika, 20, 1255

\bibitem[{{Kerdraon} \& {Delouis}(1997)}]{Kerdraon97}
{Kerdraon}, A. \& {Delouis}, J.-M. 1997, in Lecture Notes in Physics, Vol. 483,
  Coronal Physics from Radio and Space Observations, Berlin Springer Verlag,
  ed. G.~{Trottet}, 192--201

\bibitem[{{Karlick{\'y}} {et~al.}(2013){Karlick{\'y}},
  {M{\'e}sz{\'a}rosov{\'a}}, \& {Jel{\'{\i}}nek}}]{Karlicky2013}
{Karlick{\'y}}, M., {M{\'e}sz{\'a}rosov{\'a}}, H., \& {Jel{\'{\i}}nek}, P.
  2013, \aap, 550, A1

\bibitem[Klein et al.(2001)]{2001A&A...373.1073K}
Klein, K.-L., Trottet, G., Lantos, P., \& Delaboudini{\`e}re, J.-P.\ 2001, \aap, 373, 1073 

\bibitem[{{Kuijpers}(1972)}]{Kuijpers1972}
{Kuijpers}, J. 1972, in CESRA-3, Committee of European Solar Radio Astronomers,
  ed. J.~{Delannoy} \& F.~{Poumeyrol}, Vol.~3, 130

\bibitem[{{Kuijpers}(1975)}]{Kuijpers1975}
{Kuijpers}, J. 1975, \solphys, 44, 173

\bibitem[{{Kuijpers}(1980)}]{Kuijpers80}
{Kuijpers}, J. 1980, in IAU Symposium, Vol.~86, Radio Physics of the Sun, ed.
  {M.~R.~Kundu \& T.~E.~Gergely}, 341--360

\bibitem[{{Kuznetsov}(2006)}]{Kuznetsov2006}
{Kuznetsov}, A.~A. 2006, \solphys, 237, 153

\bibitem[Masuda et al.(2001)]{2001SoPh..204...55M} 
Masuda, S., Kosugi, T., \& Hudson, H.~S.\ 2001, \solphys, 204, 55 

\bibitem[Reeves \& Forbes(2005)]{2005ApJ...630.1133R} 
Reeves, K.~K., \& Forbes, T.~G.\ 2005, \apj, 630, 1133 

\bibitem[Rausche et~al.(2007)]{2007SoPh..245..327R} 
Rausche, G., Aurass, H., Mann, G., Karlick{\'y}, M., \& Vocks, C.\ 2007  \solphys,  245, 327

\bibitem[{{Rausche} {et~al.}(2008){Rausche}, {Aurass}, \& {Mann}}]{Rausche08}
{Rausche}, G., {Aurass}, H., \& {Mann}, G. 2008, Central European Astrophysical
  Bulletin, 32, 43

\bibitem[{{Slottje}(1972)}]{Slottje1972}
{Slottje}, C. 1972, \solphys, 25, 210

\bibitem[{Slottje(1981)}]{Slottje1981}
Slottje, C. 1981, Atlas of fine structures of dynamic spectra of solar type
  IV-dm and some type II radio bursts. (Dissertation, N.F.R.A Dwingeloo and
  Astronomical Institute of Utrecht)

\bibitem[Wang et al.(2001)]{2001SoPh..204..153W} 
Wang, S., Yan, Y., Zhao, R., et al.\ 2001, \solphys, 204, 153 

\bibitem[Wang et al.(2017)]{2017ApJ...848...77W} 
Wang, Z., Chen, B., \& Gary, D.~E.\ 2017, \apj, 848, 77 

\bibitem[Yan et al.(2001)]{2001ApJ...551L.115Y} 
Yan, Y., Deng, Y., Karlick{\'y}, M., et al.\ 2001, \apjl, 551, L115 

\bibitem[{{Young} {et~al.}(1961){Young}, {Spencer}, {Moreton}, \&
  {Roberts}}]{Young61}
{Young}, C.~W., {Spencer}, C.~L., {Moreton}, G.~E., \& {Roberts}, J.~A. 1961,
  \apj, 133, 243

\bibitem[{Zlobec \& Karlick{\`y}(2014)}]{zlobec2014}
Zlobec, P. \& Karlick{\`y}, M. 2014, Solar Physics, 289, 1683

\end{thebibliography}
\end{document}